\documentstyle[epsf]{laa}

\newcommand{\be}{\begin{equation}}
\newcommand{\ee}{\end{equation}}
\newcommand{\bd}{\begin{displaymath}}
\newcommand{\ed}{\end{displaymath}}
\newcommand{\bea}{\begin{eqnarray}}
\newcommand{\eea}{\end{eqnarray}}

\newcommand{\rmd}{{\rm d}}

\newcommand{\gapprox}{\;\rlap{\lower 2.5pt
             \hbox{$\sim$}}\raise 1.5pt\hbox{$>$}\;}
\newcommand{\lapprox}{\;\rlap{\lower 2.5pt
             \hbox{$\sim$}}\raise 1.5pt\hbox{$<$}\;}
\newcommand{\bfg}[1]{\setbox0=\hbox{#1}   
  \kern-.025em\copy0\kern-\wd0
  \kern.05em\copy0\kern-\wd0
  \kern-.025em\raise.0433em\box0}
\def\va{V_{\rm A}}

\begin{document}
   \thesaurus{06; 
              08.18.1; 
              02.13.1; 
              08.23.1; 
              08.13.2; 
              08.05.3 
              }
   \title{Origin of the rotation rates of single white dwarfs}

%
   \author{H.C. Spruit \inst{1}}


   \institute{ Max-Planck-Institut f\"ur Astrophysik, Postfach 1523,
               D-85740 Garching bei M\"unchen, Germany. (henk@mpa-garching.mpg.de)
               }

   \date{Received }

   \maketitle



   \begin{abstract}
I argue that the rotation of white dwarfs is not a remnant of the angular momentum of their main sequence progenitors but a result of the mass loss process on the AGB.
Weak magnetic fields, if present in stellar interiors, are likely to maintain approximately uniform rotation in stars, both on the main sequence and on the giant branches. The nearly uniform rotation of the core of the Sun is evidence for the existence of such fields. Exactly axisymmetric mass loss on the AGB from uniformly rotating stars would lead lead to white dwarfs with very long rotation periods ($>$ 10 yr). Small random non-axisymmetries ($\sim 10^{-3}$) in the mass loss process, on the other hand, add sufficient angular momentum to explain the observed rotation periods around one day. The process illustrated with a computation of the probability distribution of the rotation periods under the combined influence of random forcing
by weak nonaxisymmetries and angular momentum loss in the AGB superwind. Such asymmetries can in principle be observed by proper motion studies of the clumps in interferometric images of SiO maser emission.

\end{abstract}

\section{Introduction}
White dwarfs are observed to rotate with typical periods of a day. The main sequence
progenitors of these stars are also rotating, and it is generally assumed that the
rotation of white dwarfs is a remnant of this main sequence rotation. Arguments
involving conservation of angular momentum can be used to make this plausible (e.g.\ Perinotto 1990, Pijpers 1993). A
problem with this picture is, however, that progenitors of WD have gone through a
giant stage in which at least the envelope rotated very slowly. Thus, it is necessary 
to assume that the cores of giants remain decoupled rotationally from their envelopes 
during the entire evolution from main sequence turnoff till the formation of the WD. 
Since little is known with certainty about the processes that might redistribute 
angular momentum inside stars, this assumption can not be easily rejected.

There are, however, observational and theoretical reasons to doubt this picture. A 
strong observational argument is the internal rotation of the Sun. The most recent 
helioseismological measurements (Elsworth et al. 1995, Kosovichev et al. 1997, Corbard et al. 1997) show that the rotation below the convection zone is esentially uniform, with measured degrees of differential rotation well below the 30\%. seen at the surface. The known hydrodynamic angular momentum transport processes, even with rather optimistic estimates of their efficiency, leave the Sun with a much too rapidly rotating interior (Spruit et al 1983).
A new hydrodynamic mechanism recently studied in some detail is friction by internal
gravity waves excited by the convection zone (Press 1981, Spruit 1987, Zahn 1990,
Schatzman 1993). Realistic calculations of this process appear to be difficult, but 
estimates indicate that it can be more effective than the other hydrodynamic processes 
(Zahn et al. 1997).

Magnetic fields, on the other hand, have long been known to be very efficient at 
transporting angular momentum. The torques exerted by magnetic fields become significant 
already at very low field strengths. For the Sun, for example, a field of less than 10G 
can provide sufficient torque to maintain the observed uniformity of rotation. A number 
of mechanisms can provide such weak fields, for example a fossil field (remnant of the 
star formation process) or a dynamo-like process operating on (a small remnant of) the 
differential rotation of the core.

In this paper, I develop the consequences of assuming that the cores of  giants do, 
in fact, corotate approximately with their envelopes. After discussing the observational 
evidence on WD rotation rates I develop theoretical arguments for the existence of 
effective coupling between the core and the envelope. This predicts very slowly rotating 
cores in the giant progenitor of a single WD. The rotation of single white dwarfs must 
then be explained by other processes.

The same applies to neutron stars born in red giants. The observed pulsar rotation
periods of the order of a second are much shorter (by a factor $10^3$ or so) than
expected if they formed in approximately uniformly rotating giants, and with our
assumption of strong coupling of the core another mechanism also has to be found to
explain the rotation of pulsars. The processes differ somewhat for white dwarfs and 
neutron stars. The arguments for the neutron star case are developed in a companion 
paper (Spruit and Phinney, 1998). There, we show that the kicks with which neutron 
stars are born (as inferred from their transverse velocities) also impart angular 
momentum at amounts sufficient to explain the rotation of most pulsars.

To explain the typical rotation rates periods of single white dwarfs (of which
only about 30 have measured rotation rates), I show in Sect.\ \ref{nonax} that small 
asymmetries in the mass loss processes during the last phases of evolution on the AGB are sufficient to explain the observed rotation rates. These asymmetries act as a random forcing through which angular momentum accumulates in the envelope. A balance results between this random forcing and the loss of angular momentum by the wind. The evolution of the angular momentum as the mass loss proceeds turns out to be mathematically the same as that of the velocity of a particle experiencing Brownian motion in a gas, and can be described by a Fokker-Planck equation. Solutions of this equation (Sect.\ \ref{solu}) show that probability distribution of the angular momentum is close to a Maxwellian. The mean angular momentum decreases as the square root of the envelope mass remaining. Current observational evidence relating to the asymmetries needed in this picture is discussed in Sect.\ \ref{discu}.

\subsection{Rotation speeds of AGB cores}
Starting with a rapidly rotating main sequence star, and assuming uniform rotation
during the expansion to the giant stage, we can estimate the rotation periods to
be expected for white dwarfs evolving from single stars. An early type main
sequence (MS) star, rotating near its maximum speed (of the order 400 km/s), and
expanding without angular momentum loss onto the asymptotic giant branch (AGB), has a 
rotation period $P_{\rm G}=2\pi/(GM/R_{\rm G}^3)^{1/2} (R_{\rm G}/R_{\rm MS})^{1/2} 
\sim 10$yr for $R_{\rm G}\sim 1$AU (except for a modest difference in gyration radius neglected here). Most early type MS stars rotate significantly slower, so that periods of the order 30-100 yr would be expected for the AGB descendants of early 
type stars. Some of the observed white dwarfs must have descended from solar type stars 
(F-G), which have periods of the order 30d at the end of their main sequence life. The 
AGB progenitors of these WD would rotate 100 times slower, with periods of the order 
of a thousand years.

If the small amount of envelope mass is ignored which settles back onto the core during 
post-AGB evolution (more about this in Sect.\ \ref{axisy}), these rotation periods would 
also be inherited by the WDs formed. While there are a few magnetic white dwarfs with 
inferred rotation periods of at least a century, most WDs for which periods are known 
rotate much faster. We evidently need another mechanism to explain the rotation of typical 
single WDs. Before entering the discussion of possible mechanisms, I briefly review the 
observational evidence on WD rotation.

\section{White dwarf rotation periods}
\label{wdrot}
\begin{figure}[t]
\mbox{}\epsfysize6cm\epsfbox{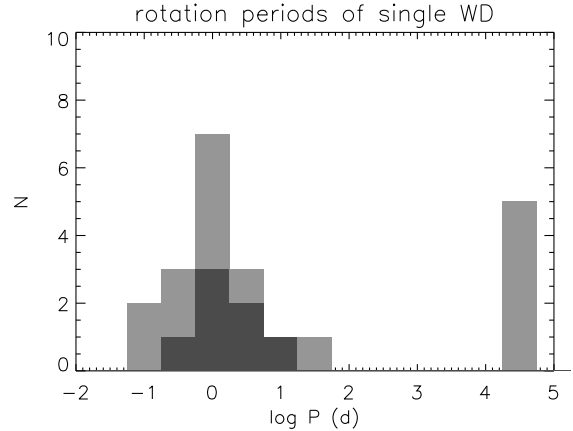}\hfill\mbox{}
\caption{\label{hist}
Rotation periods of isolated white dwarfs. Dark: oscillating WD
(asteroseismological periods), light: magnetic WD. From Schmidt
and Norsworthy 1991, Putney 1996, and refs in Table 1. The peak at the right
represents the lower limits for the 5 stars whose rotation period is larger than
about a century.}
\end{figure}

Three methods exist for measuring the rotation rates of white dwarfs. The
largest number of determinations comes from spectral and/or polarization
variations in magnetic WDs. Figure \ref{hist} shows these rotation periods, as
compiled by Schmidt and Norsworthy (1991, see also Schmidt and Smith, 1995).
Added to this compilation was G 158-45 (Putney, 1996) with a period of 4.44 hr.
A second group of determinations comes from ZZ Ceti and other oscillating stars.
In a number of these, sufficiently detailed observations exist to identify the
oscillation modes, allowing determination the period splittings due to rotation.
The rotation periods for 7 oscillating white dwarfs collected from the
literature (Table 1) are also shown in Fig \ref{hist}. The spike in the figure
at $3\,10^4$d represents the (magnetic) stars whose periods are inferred to be
longer than a century, on the basis of the absence of variations in the
polarization on time scales of decades. These stars were put at their approximate
lower limits of 100 yrs.

The widths of the narrow NLTE line cores have been used to set limits on
rotation velocities of stars for which the magnetic and seismological methods
can not be used (Wesemael et al. 1980, Koester and Herrero 1988, Koester et al., 1998, in prep.). The detection limit, apparently around 20 km/s, is not sensitive enough to determine the
rotation of stars in the $\sim 1$d main peak in Fig.\ 1, but may be useful in
setting limits on the number of rapidly rotating ($P<1$hr) stars. Reid (1996, his
Sect.\ 3.1) and Heber et al. (1997) infer upper limits from 8 to 40 km/s from Keck spectra of some 25 single white dwarfs.

The distribution of periods in the main hump around 1d looks the same for the
magnetic and the oscillating stars, given the limited statistics. Very long
periods are absent from the sample of seismologically determined periods, but
this may be due to observational limitations. No stars have had their
oscillations followed long enough to detect period splittings of a decade.
One concludes that with the (limited) data available, there does not seem to be a
significant difference in the distribution of rotation rates of magnetic
($B\gapprox 10^5$G) and nonmagnetic ($B\lapprox 10^4$G) stars. There may, however, be other differences between the magnetic and nonmagnetic WD, apart from the field
strength. Sion et al. (1988) and Liebert (1995) for example, argue that the magnetic
stars are more massive than the nonmagnetic ones, and derive from more massive
progenitors.

\begin{table}[ht]
\begin{tabular}{lcl}
object & P & ref \\
PG 1159$-$035 & 1.38d & Winget et al. 1991 \\
PG 2131+066 & 5.07hr & Kawaler et al. 1995 \\
PG 0122+200 & 1.61d & O'Brien et al. 1996 \\
GD 358 & 1.38d & Winget et al. 1994 \\
GD 165 & 4.2d & Bergeron et al. 1993 \\
GD 154 & 2.3d & Pfeiffer et al. 1995 \\
G226$-$29 & 8.9h & Kepler et al. 1995 \\
\end{tabular}
\caption{Asteroseismologically determined WD rotation periods}
\end{table}

\section{Angular momentum transport processes}

\subsection{internal gravity waves}
The kinetic energy of convective motions in the envelope appears as a source of
pressure fluctuations at the boundary between the envelope and the core. These
fluctuations propagate as internal gravity waves through the core. The waves
carry angular momentum and their dissipation therefore transmits torques. Assume
that prograde and retrograde waves are excited with the same amplitude. This is
a good approximation if the rotation is slow, such that the effect of Coriolis
forces on the wave generation process is small. As the waves propagate into the
core, they  conserve their angular momentum (or wave action) until dissipation
becomes important. If the dissipation of prograde and retrograde waves is the
same, no net angular momentum transport takes place. Prograde (retrograde) waves
propagating into a medium of increasing (decreasing) rotation speed, however,
meet critical layers (where the rotation rate equals the horizontal component of
the phase speed), and are much more effectively absorbed there (see the
discussions in Goldreich and Nicholson 1989, Zahn et al. 1997). Due to this
asymmetry, there is a net angular momentum transport which tends to reduce the
differential rotation. In effect, the internal gravity wave field is a source of
friction. Zahn et al. find that for the Sun, the time scale for synchronization
between core and envelope due to this friction is of the order $10^7$ yr. Since
this is of the order of evolution time scale to the giant branch, the process
could be significant in maintaining synchronization during core contraction. Detailed evolution calculations including this process by Talon and Zahn (1998), however, still yield too large internal rotation rates for the present Sun.

\subsection{magnetic torques}
\label{mt}
Magnetic torques are transmitted by the stress component $B_rB_\phi/(4\pi)$.
Approx\-imat\-ing this as constant over a spherical surface, the angular momentum
balance is
\be I\dot\Omega={2\over 3}B_rB_\phi r^3. \label{ab}\ee
The synchronization time scale between a core of radius $r=10^{10}R_{10}$cm, mass
$M=mM_\odot$ and moment of inertia $I=k^2MR^2$ rotating at a rate $\Omega=10^{-6}
\Omega_{-6}$s$^{-1}$ is then
\be  \tau={3\over2}{k^2M\Omega\over
B_rB_\phi R}=3\,10^{8} (B_rB_\phi)^{-1} m r^{-1}_{10} \Omega_{-6} \quad{\rm yr},
\ee
where $k$ is the gyration radius. At the Sun's current rotation rate $\Omega_{-6}=3$, the synchronization time scale is less than the age of the Sun if $(B_rB_\phi)^{1/2}>3$G. If on its way to the
giant branch the core of the Sun were to contract to a radius of $10^9$cm while
conserving angular momentum, it would rotate at rate $\Omega_{-6}\sim 300$. To
maintain corotation on the $10^7$ yr contraction time scale, a field strength
$(B_rB_\phi)^{1/2}=300$G is sufficient, at this rotation rate.

If the azimuthal and radial field components are of similar magnitude, (\ref{ab}) can be written in terms of the magnetic energy $E_B=B^2/(8\pi)(4\pi/3) r^3$, and rotational energy $E_\Omega={1\over 2}I\Omega^2$ as
\be E_B={1\over 2}E_\Omega{1\over \tau\Omega}.\label{tob}\ee
If the $B_r$ and $B_\phi$ are not comparable, (\ref{tob}) is only the minimum magnetic energy required. Nevertheless, since $\tau\Omega$ is typically such a large number,  (\ref{tob}) shows that corotation can be maintained by a magnetic energy which is {\it small fraction of the rotational energy} of the star, for spindown on a time scale long compared with the rotation period.

\subsection{Winding up of field lines}
Are such magnetic field strengths plausible? If the fields of the magnetic A stars
are fossil (which unfortunately is still unclear), sufficiently strong fields might
also exist in the cores of solar type stars. Even if the initial fields (on the ZAMS) are lower than these values, however, differential rotation will increase the field strength quickly to values that have an effect on rotation. Whether initially present in the star or developing later by core contraction, differential rotation winds up the field lines, increasing the field strength. This problem has
been studied in various forms since the '50s. Winding up of an axially symmetric poloidal field into a predominantly azimuthal field by differential rotation produces an opposing torque that is linear in the number of differential turns made,
as in a harmonic oscillator. The result is an oscillation of alternate winding up and unwinding at a period given by the Alfv\'en travel time through the star (Mestel, 1953), where the Alfv\'en speed is that of the poloidal field (which is unaffected by the winding-up). Since Alfv\'en waves travel decoupled from each other, each on its own magnetic surface, the oscillation period is different on each magnetic surface. The oscillations on these surfaces therefore get out of phase after a few oscillations, and the length scale across the surfaces decrease as $t^{-1}$. In a finite time,  dissipative processes across the surfaces become important, and the oscillation damps out by phase mixing (Spruit 1987, Charbonneau and MacGregor 1993, Sakurai et al. 1995).  The net effect of the process is that the component of differential rotation along the field lines is damped out, on a finite time scale, and this can happen with an initial field that is much weaker than estimate (\ref{tob}).

\subsection{Magnetic shear instability}
Another possibility is that a turbulent field is generated by the same magnetically mediated shear instability that has been shown to operate effectively in accretion disks (Hawley et al. 1995, Matsumoto and Tajima 1995, Brandenburg et al.\ 1995). The conditions for magnetic shear instability to exist in a star have already been studied in detail by Acheson (1978, 1979) though the proper interpretation of this instability (Balbus and Hawley, 1992) was not clear at the time (see, however, Fricke, 1969). In the context of stellar interiors, it has been studied again recently by Kato (1992), Balbus and Hawley (1994) and Urpin (1996). Wherever this instability exists it will lead to very rapid growth (on the differential rotation time scale) of a turbulent magnetic field, which then acts on the differential rotation like an effective viscosity.

Acheson's (1978) analysis of the instability conditions includes (unlike the more recent works) the effects of thermal and magnetic diffusion and of viscosity. The inclusion of thermal diffusion is especially important since it makes the instability appear under much wider conditions. This is seen from Acheson's condition (7.27, a special case of his more general condition), which is equivalent to
\be
-2q-{\va^2\over\Omega^2r^2}\left({r\over \gamma H}-2\right)F>{\eta\over\kappa} {\gamma N^2\over\Omega^2}, \qquad ({\rm for~instability}),\label{ach}
\ee
where
\be q={\rm d}\ln\Omega/{\rmd\ln r},\qquad F={\rm d}\ln B_\phi/{\rmd\ln r},\ee
$N$ is the buoyance frequency, $\eta$ and $\kappa$ the magnetic and thermal diffusivities, and $H$ the pressure scale height. This condition holds for low viscosity ($\nu/\eta\ll 1$), for an azimuthal field $B_\phi$ at the equator of the star. The first term on the left hand side represents the magnetic shear instability, the second term Parker instability (magnetic buoyancy instability).
For weak fields, this second term is negligible. The right hand side shows the stabilizing effect of the stratification, which, however, is partially undone by thermal diffusion (for adiabatic perturbations, the factor $\eta/\kappa$ would be replaced by unity). Since photons diffuse so much more effective than the magnetic field, the instability is present much more widely than in an adiabatic treatment. The instability, however, is able to grow only on length scales sufficiently small that thermal diffusion is important. This somewhat limits its effectiveness, and it may be that the effective viscosity it produces is not much larger than the viscosity produced by hydrodynamic shear instabilities (Zahn, 1974) under the same conditions. These questions could, in principle, be readily addressed by an appropriate numerical simulation.

Because magnetic fields are so effective at transmitting torques, already at low field strengths, differential rotation can survive over a large number of rotations only in regions where the radial field component is very small. In order
to allow the core in a giant to rotate substantially faster than its envelope, one must find a reason why it could have been so accurately `shielded' magnetically, over the entire life of the star on the giant branch.

While the arguments given here do not constitute a proof, I feel they are sufficiently strong that approximately uniform rotation (modulo a factor of a few) is a reasonable hypothesis, compared with the traditional assumption in which the core of a giant rotates $10^4$--$10^5$ times faster than its envelope.

\section{Mass and angular momentum loss on the AGB}
A large fraction of the star's mass is lost in the last phases of evolution on the AGB (e.g.\ Habing, 1990). Most is ejected in the form of a superwind ($\sim 10^{-4} M_\odot{\rm yr}^{-1}$) lasting on the order of $10^4$yr (e.g. Vassiliadis and Wood, 1993). It is believed to be driven by pulsational instability and radiation pressure on dust (Fleischer at al. 1992, Sedlmayr and Carsten 1995, H\"ofner and Dorfi, 1997), or possibly by sound waves (e.g. Pijpers and Hearn, 1989). The mass loss is probably not steady because the stellar pulsation is an important part of the driving. Also, dust formation in the expanding flow is an unstable process (H\"ofner and Dorfi, 1997). Thus the envelope is probably ejected in the form of a (large) number of light shells. The mass loss is also believed to be modulated on longer time scales by the thermal pulses of the AGB star.

A small fraction of the star's envelope (on the order $10^{-4}M_\odot$) settles back onto the core after the superwind ceases. Most of the angular momentum is lost together with the mass of the envelope, but because of the large size of the envelope, even the small amount of mass remaining might conceivably contain enough angular momentum to form a significantly rotating white dwarf. Thus we need to look in some detail at the angular momentum balance of the mass losing AGB envelope. First, I show that if the superwind is axially symmetric and has the specific angular momentum of the stellar photosphere from which it is ejected, the angular momentum remaining after envelope ejection is {\it far} too small to produce a significantly rotating white dwarf.

\subsection{Axially symmetric mass loss}
\label{axisy}
If the mass is ejected from the stellar photosphere in axisymmetric fashion, taking with it the angular momentum it had in the photosphere, the net angular momentum loss by the wind is
\be\dot J={\textstyle{2\over 3}}\dot M R_*^2\Omega_*,\ee
where $\dot M$ is the mass loss rate, $R_*$, $\Omega_*$ the photospheric radius and rotation rate of the envelope. The factor 2/3 is due to the variation of specific angular momentum over the surface. Since the envelope is convective, it is a good approximation to assume that it rotates uniformly. Because of the very large radius of the envelope, the core contributes very little to the star's moment of inertia, even if the envelope mass is quite small. By angular momentum conservation the star's angular momentum $J_*$ varies as
\be \dot J_*={\textstyle{2\over 3}}\dot M_* R_*^2\Omega_*,\ee
where $M_*$ is the star's mass. With uniform rotation, $J_*=k^2M_*\Omega R_*^2$, where $k$ is the radius of gyration, hence
\be \dot J_*={2\dot M_*\over3 k^2M_*} J_*. \label{dotj}\ee
In stars, $k^2<0.4$, so that the angular momentum of the star decreases more rapidly than its mass. This is because the specific angular momentum of the mass leaving the star is higher than the average specific angular momentum of the star (by a factor $2/3k^2$). Since the envelope mass varies strongly, the gyration radius can not be taken as constant. The total moment of inertia of the star can be written as the sum of core and envelope contributions:
\be I_*=k^2M_*R_*^2=\int_{\rm core}^{ }\rho\varpi^2 {\rm d}^3{\bf r}+\int_{\rm envelope}^{ } ...= I_{\rm c}+I_{\rm e},\ee
where $\varpi$ is the distance to to rotation axis, and
\be I_{\rm c}=k_{\rm c}^2M_{\rm c}R_{\rm c}^2,\qquad I_{\rm e}=k_{\rm e}^2M_{\rm e}R_*^2.\ee
If the envelope contains most of the stellar mass, $k_{\rm e}^2$ is approximately that of a polytrope of index 1.5, $k_{\rm e}^2\approx 0.2$. %
For the estimates below I assume this value.
For a degenerate core of mass $\sim 0.6$, $k_{\rm c}^2$ is of the order 0.19. The gyration radius of the star as a whole is then
\be k_*^2=I/(M_*R_*^2)=k^2_{\rm c}{M_{\rm c}\over M_*}{R_{\rm c}^2\over R_*^2}+k_{\rm
e}^2{M_{\rm e}\over M_*}.\ee For $R_*\sim 10^{13}$, $R_{\rm c}\sim 10^9$, the first term is
negligible for envelope masses larger than $\sim 10^{-8}M_\odot$, so that \be k_*^2\approx
k_{\rm e}^2{M_{\rm e}\over M_*}.\ee With (\ref{dotj}) this yields
\be
\dot J_*/J_*=m{\dot M_*\over M_{\rm e}}.\qquad(M_{\rm e}\gapprox 10^{-8}) ,\label{dotj2}
\ee
where
\be m={2\over 3k_{\rm e}^2}\approx 3.3. \ee

Since the core mass is essentially constant during the mass loss, we have $\dot M\approx \dot M_{\rm e}$. Eq. (\ref{dotj2}) can be integrated to yield
\be {J_*\over J_0}=\left({M_{\rm e}\over M_{{\rm e}0}}\right)^m, \label{jm}\ee
where $J_0$ and $M_{{\rm e}0}$ are the initial angular momentum and envelope
mass. The steep dependence on $M_{\rm e}$ implies that only a small fraction of $J_0$ is retained. An upper limit on the final rotation rate is obtained by assuming the AGB star to rotate critically, $\Omega_*=(GM/R_*^3)^{1/2}$.
The rotation rate of the post-AGB core then becomes
\be  \Omega_{\rm f}/\Omega_0={k_*^2M_*\over k_{\rm f}^2M_{\rm f}}{R_*^2\over R_{\rm f}^2} \left({GM_*\over R_*^3}\right)^{1/2} \left({M_{\rm
ef}\over M_{{\rm e}0}}\right)^{3.3}, \ee where indices $_{\rm f}$ and $_*$ denote the
post-AGB core and the AGB star, respectively. If at the end of the superwind phase an
envelope mass of not more than $10^{-3}M_{{\rm e}0}$ is left, we get a final rotation period of at least a year.

The effect depends rather critically on the index $m$ in (\ref{jm}). If the wind corotates with the star out to some radius $R>R_*$, for example because of an atmospheric magnetic field, the specific angular momentum in the wind is increased by the factor $f=(R/R_*)^2$, and the index $m$ would become
\be m={2f\over 3k_{\rm e}^2}. \label{mli}\ee
 Magnetic fields are known to exist in Mira envelopes from the circular polarization of the SiO
masers (Barvainis et al. 1987, Kemball and Diamond 1997). The values of the field derived are
uncertain since they depend on the degree of saturation of the masers (Nedoluha and Watson 1994).
A strength of a few tenths of a Gauss, however, would already cause significant additional angular
momentum loss by the wind.

The conclusion is that even a maximally rotating AGB star, with its huge amount of angular momentum, will produce only a nearly non-rotating white dwarf if mass loss is axisymmetric. This results from the fact that almost all the envelope is lost, combined with the higher than average specific angular momentum taken away by the mass lost. Physically, as mass is lost from the photosphere, the envelope expands, causing spindown by angular momentum conservation.

Let me summarize the assumptions made in arriving at this, perhaps surprising, conclusion. The first is that core of the AGB star corotates approximately with the envelope when the phase of rapid mass loss sets in. The others are the rather minimal assumptions that the (convective) envelope rotates approximately uniformly, and that the mass lost in the wind carries at least the specific angular momentum of the photosphere of the star.

\section{Slightly nonaxisymmetric mass loss}
\label{nonax}
The angular momentum evolution of the star is altered dramatically if even a small
amount of non-axisymmetry is allowed in the mass ejection process. If a shell is ejected aspherically, it generally carries a net momentum, and the direction of this                  momentum vector in general need not pass exactly through the center of mass. It is conceivable, for example, that the dust-formation instability found in spherically symmetric numerical simulations of the ejection process actually is non-axisymmetric, so that the forces exerted are not evenly distributed over the surface. In this way, the ejection process adds a small amount of angular momentum (`kick') to the star.

Suppose now that a large number of shells are ejected, adding small amounts of angular momentum in random directions. Since the simulations indicate that the shells are ejected with periods $\delta t$ of the order of the oscillation period of the star (on the order of a year) while the duration of the superwind phase is of the order of $10^4$yr, there are on the order of $10^4$ kicks, each associated with an ejected mass $\delta M$ on the order of $10^{-4}M_\odot$. The maximum amount of angular momentum such a kick can impart is $\delta MR_*^2 v_{\rm e}$, where $v_{\rm e}$ is the ejection velocity, observed to be in the range 5--50 km/s. This maximum applies when the mass is ejected tangentially to the surface of the star. This is of course quite unrealistic, and one expects the angular momentum imparted to be only a small fraction of this:
\be \delta J=\epsilon\, \delta MR_*^2 v_{\rm e}, \label{dj}\ee
where $\epsilon$ is a small number. In the following, I estimate how large this number must be to explain the observed rotation periods.

The evolution of the star's angular momentum vector is obtained by adding the forcing by kicks to (\ref{dotj2}):
\be
\dot {\bf J}=-m{\vert\dot M_{\rm e}\vert\over M_{\rm e}}{\bf J}+{\bf A}(t), \ee
where ${\bf A}$ is a random fluctuating vector with time step $\delta t$ and amplitude $\delta J$. This equation (Langevin's equation) is the same as that governing the Brownian motion of particles in a gas. Following the standard treatment in statistical physics (e.g. Becker, 1978) we can take the continuum limit, in which the time step is infinitesimal, and derive a Fokker-Planck equation for the probability distribution $f({\bf J},t)$ of obtaining an angular momentum ${\bf J}$ after time $t$. Leaving out this derivation, the result is
\be {\partial f/\partial t}= \beta \nabla_J\cdot({\bf J}f+D\nabla_J f), \label{fp3}\ee
where
\be \beta=m{\vert\dot M_{\rm e}\vert\over M_{\rm e}},\label{beta}\ee
is the `braking rate', $\nabla_J$ is the gradient in ${\bf J}$-space, and  $D$ the diffusion coefficient in ${\bf J}$-space
\be D={\textstyle{1\over 3}}(\delta J)^2/\delta t .\label{D}\ee
The main difference with respect to standard Brownian motion is that the coefficients $\beta,D$ in the present case are functions of time.

If the kicks are random in direction, and the star initially non-rotating, the probability distribution $f$ is isotropic in ${\bf J}$-space, $f=f(J)$.  Writing
\be F=J^2f,\ee
Eq. (\ref{fp3}) can then be written as
\be \partial_tF=\partial_J[\beta JF+D(\partial_JF-2F/J)]. \label{fp1} \ee
If $\beta$ and $D$ are constant, as they are in the case of Brownian motion, the asymptotic solution $F_\infty$ for large $t$ is that for which the bracket on the RHS vanishes. This yields
\be F_\infty\sim J^2\exp(-{1\over 2}{\beta\over D}J^2), \label{max}\ee
i.e. a Maxwellian distribution peaking at $J=(2D/\beta)^{1/2}$. In our case, $\beta$ varies significantly with time, because the envelope mass varies strongly. We should therefore do not expect the distribution function to be a Maxwellian.

Before entering into more detailed calculations an estimate of the orders of magnitude to be expected can be made by making a quasi-stationary approximation to
(\ref{fp3}). The distribution $F$ is then given approximately by (\ref{max}). The typical angular momentum to be expected at the end of the mass loss, when the envelope mass left is $M_{\rm e}\ll M_{{\rm e}0}$ is then, with (\ref{beta}), (\ref{D}):
\be J\approx\delta J(nk_{\rm e}^2 M_{\rm e}/M_{{\rm e}0})^{1/2},\ee
where
\be n=(M_{{\rm e}0}-M_{\rm e})/(\dot M_{\rm e}\delta t)\ee
is the number of kicks experienced. With (\ref{dj}):
\be  J\approx\epsilon M_{{\rm e}0}R_* v_{\rm e} \left({M_{\rm e}\over mnM_{{\rm e}0}}\right)^{1/2}, \label{jn}\ee
where $m=2/(3 k_{\rm e}^2) $ for angular momentum loss at the photospheric value
($f=1$ in \ref{mli}). The expected rotation rate of the white dwarf, with gyration radius
$k_{\rm w}$, mass $M_{\rm w}$ and radius $R_{\rm w}$ is
then $\Omega_w=J/(k_{\rm w}^2M_{\rm w}R_{\rm w}^2)$. Assuming a final envelope mass of $10^{-4}M_\odot$, initial envelope mass of $2M_\odot$, and $M_{\rm w}=0.6M_\odot$, this yields
\be \Omega_w\approx \epsilon 5 n^{-1/2} {\rm s}^{-1}.\ee
If shells are ejected every two years or so, we have $n\approx 10^4$. A rotation period of 1d is then obtained for $\epsilon\approx 10^{-3}$.

As long as it is not known how the relevant details of the ejection process take place, it is hard to argue whether an asymmetry of the order $10^{-3}$ is realistic or not, but a number as small as this would not seem too demanding. The reason why such small asymmetries are sufficient, even when their effect is further reduced by random superposition (the factor $n^{-1/2}$ in \ref{jn}), is the very large lever arm on which the kicks act. A star on the AGB is so large compared with the final white dwarf that a very precisely axisymmetric mass loss would be needed to {\it avoid} introducing the small amount of angular momentum that is sufficient to produce white dwarfs with periods of a day.

\subsection{Distribution of rotation rates resulting from random kicks}
\label{solu}
The coefficient $\beta$ varies by a factor $10^4$ as the envelope mass is reduced from its initial value to a representative post-AGB value of the order $10^{-4}M_\odot$. To take this into account, I solve Eq. (\ref{fp1}) numerically. I use a second order, implicit time step and centered differences in the $J$-coordinate (Crank-Nicholson scheme).

\begin{figure}[t]
\mbox{}\epsfysize6cm\epsfbox{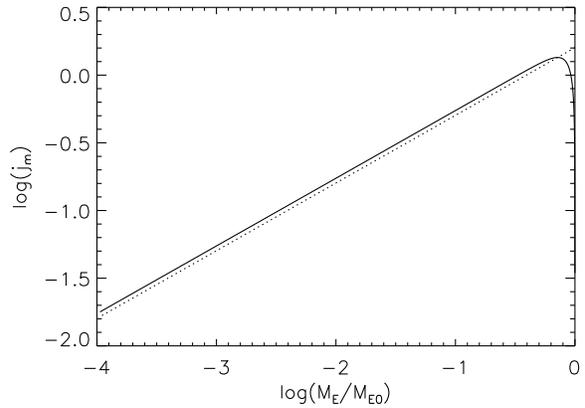}\hfill\mbox{}
\caption{\label{jmt}
The mean angular momentum $j_{\rm m}$ as a function of the
remaining envelope mass $M_{\rm e}$, for angular momentum induced by random non-axisymmetries in the superwind mass loss. Solid: solution of Eq. (33). Dotted: stationary approximation, Eq. (34). Evolution is from right to left}
\end{figure}

As angular momentum coordinate I use the dimensionless variable $j$, defined by
\be J=j \epsilon M_{{\rm e}0} R_* v_{\rm e}(mn)^{-1/2}. \ee
Let
\be g(j)=j^3f(j)=\rmd N/\rmd \ln j\ee
be the probability distribution per unit of $\ln j$, and for time coordinate use
\be \tau=-\ln(M_{\rm e}/M_{{\rm e}0}). \ee
Then Eq.~(\ref{fp1}) can be written as
\be \partial_\tau g=j\partial_j[mg+e^{-\tau}j^2\partial_j(g/j^3)]. \label{fpn} \ee
The integration is from $\tau=0$ to $\tau_{\rm f}=-\ln(M_{{\rm e}f}/M_{{\rm e}0})$, where
$M_{{\rm e}f}$ is the envelope mass at which the mass loss ends. Apart from  $M_{{\rm e}f}$
the only parameter in the problem is the angular momentum loss index $m=2f/(3k_{\rm e}^2)$ (cf. Eq. \ref{mli}). The evolution for $m=3.3$ is given in Fig.\ \ref{jmt}, which shows the mean $j_{\rm m}$ of the probability distribution $g(j,M_{\rm e})$ as a function of the remaining envelope mass.

If the evolution is sufficiently `slow', one expects the solution to be close to the Maxwellian stationary solution, obtained by setting the square bracket in (\ref{fpn}) equal to zero. This stationary distribution has mean dimensionless angular momentum
\be j_s=2\sqrt{2/\pi}(M_{\rm e}/M_{{\rm e}0})^{1/2},\label{jinf} \ee
and is shown for comparison in Fig.\ \ref{jmt}. The stationary approximation actually turns out to be quite good, except in the initial phase of the evolution.
The white dwarf rotation rate corresponding to (\ref{jinf}) is given by:
\be
\Omega_w=\epsilon\, 2\sqrt{2/\pi}{M_{{\rm e}0}\over k_{\rm c}^2 M_{\rm w}}{R_*\over R_{\rm w}} {v_{\rm e}\over R_{\rm w}}
({M_{\rm e}\over mnM_{{\rm e}0}})^{1/2}.\label{wdr}
\ee
Thus, the predicted white dwarf rotation rate decreases as the square root of the mass remaining in the envelope at the time when mass loss ceases.

\subsection{Comparison with observed distribution}
\begin{figure}[t]
\mbox{}\epsfysize6cm\epsfbox{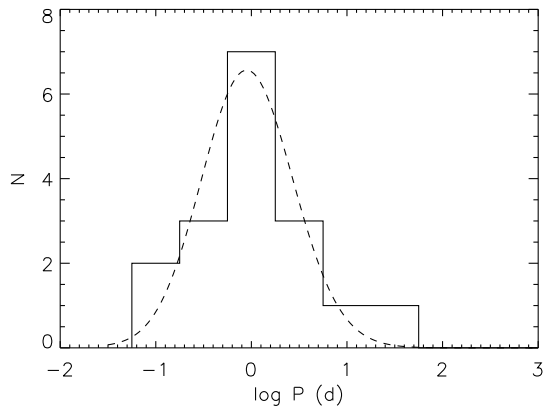}\hfill\mbox{}
\caption{\label{medist}
Predicted distribution of rotation periods (solid) for asymmetry parameter $\epsilon=10^{-3}$, and a log-normal spread in final envelope
mass from $2\,10^{-5}$ to $5\,10^{-4}M_\odot$. Histogram: observed distribution from Fig.\  1}
\end{figure}

By adjusting either the asymmetry parameter $\epsilon$ or the final envelope mass $M_{{\rm e}f}$, the maximum of the distribution can be made to agree with the observations. This distribution is close to a Maxwellian, and its width is too narrow compared with the observations, which spread by a factor 20 or so. The factors influencing the mean rotation rate (\ref{wdr}) most are the asymmetry parameter $\epsilon$ and the remaining envelope mass $M_{{\rm e}0}$. Both might depend on systematic factors like the initial stellar mass. A random variable could be the phase in the thermal pulse cycle at which the superwind takes place, which is known to have an effect on the post-AGB evolution (Sch\"onberner 1990, Vassiliadis and Wood 1993). Lacking a sufficiently detailed theory for the superwind, it is hard to guess how the asymmetry parameter might depend on such
variables. Values of the remaining envelope mass, on the other hand, have been inferred for oscillating WD and post-AGB stars by comparisons with theoretical models. Clemens (1994) finds a hydrogen envelope masses of about $10^{-4}M_\odot$.
In the helium (DB) white dwarfs and their possible progenitors the PG 1159 stars,
only a helium envelope (with inferred masses of the order $10^{-3}M_\odot$, cf. Dehner and Kawaler 1995) is left. Bl\"ocker and Sch\"onberner determine a hydrogen envelope mass of $3\,10^{-4}M_\odot$ for FG Sge. It seems reasonable to assume that a certain spread in $M_{{\rm e}f}$ is present. This could be due, for example, to random variations in moment at which pulsation ceases. To fit the observed distribution with such a spread, I assume a log-normal distribution of the parameter
$M_{{\rm e}f}$, with peak at $\bar M_{{\rm e}f}=10^{-4}M_\odot$ and (1/$e$-) width from $2\,10^{-5}$ to $5\,10^{-4}$. The asymmetry parameter is assumed to
be $\epsilon=10^{-3}$. The resulting period distribution is compared with the observations in Fig.\  \ref{medist}. The agreement with the observations is not a test of the theory developed here (since both the width and the mean have been fitted), but comparison shows that a spread in envelope mass of a factor 5 on both sides of the mean is  sufficient to explain the observed width.

\subsection{Binarity and common envelope}
Alternatives to the picture sketched may be envisaged, in which the rotating white dwarfs are in one way or another the result of binarity. The following is a brief discussion of such scenarios.

The observed periods, $0.1{\rm d}<P<10{\rm d}$ are suggestive of the orbital periods of close binaries, as has been noted by Schmidt et al. (1986), who suggest the possibility that AM Her stars might be the progenitors of the rotating white dwarfs. Those must then have somehow lost their companions, perhaps through the mass transfer. Current understanding of the evolution of CVs does not favor this possibility, since it predicts that the secondaries can not transfer all of their mass within a Hubble time. This is because the angular momentum loss slows down dramatically once the secondary has been reduced to a small degenerate dwarf (e.g. Verbunt 1996, Kolb 1993).

A second possibility that suggests itself is that of a binary companion absorbed in a common envelope (CE)
process. Two different outcomes of such a CE are possible. One is that the envelope is ejected, by the orbital
energy released, before the secondary has spiraled in completely. In the other, the secondary spirals in
completely and merges with the primary. The first case leaves a detached system (such as V471 Tau) which then evolves into a CV by magnetic braking. Theory and numerical simulations (for reviews see Taam 1995, Livio 1996) predict that this case happens if the secondary is massive enough and the density gradient in the inner parts of the giant are not too steep. If these conditions are not met, the secondary is predicted to dissolve completely, transferring all its mass to the giant envelope. A significant fraction of common envelope systems may actually experience this fate.

The high incidence of elongated or bipolar structures in planetary nebulae (PN) and objects believed to be in transit from an AGB star to a PN suggests that a large fraction of PN involve some form of common envelope evolution
(e.g.\ Han et al.\ 1995). Detailed hydrodynamical simulations have been made to reproduce the morphology of these nebulae (Icke et al. 1992, Frank and Mellema 1994). The results show that in the initial phases of the radiation driven
nebular expansion there must have been a thick disk-like structure inhibiting fast outflow in the plane of the disk, leaving a structure of two rapidly
expanding lobes and a more slowly expanding ring. In the CE interpretation, the disk contains the mass ejected in the spiral-in process.

If the secondary is small, the energy released as it spirals in is insufficient to eject the entire envelope of the primary. The net effect in this case is that both the mass and the angular momentum of the secondary are added to the envelope of the primary. The envelope remaining on the primary after the CE would then contain a large amount of angular momentum, even if the companion absorbed is small. Would this suffice to produce a rotating white dwarf? If our basic assumption of approximately uniform rotation is valid, the answer is negative. This follows from the analysis of Sect.\ \ref{axisy}, where I have shown that even a maximally rotating AGB star leaves a core rotating with a period of at least 10 years.

This answer applies as long as there is still a significant amount of mass left in the envelope after the common envelope process ($0.1M\odot$, say), and mass loss then continues like in normal AGB stars. If any significant amount of mass is left in the form of a convective envelope after the CE, the results from Sect.\ \ref{axisy} predict that the result will be a very slowly rotating white dwarf.

The consequence of the above is that a rapidly rotating white dwarf by CE evolution is to be expected only if the final dissolution of the companion coincides rather precisely with the ejection of the last bits of envelope. Barring possible surprises concerning late phases of CE evolution, the details of which are not well known, this situation would appear to be a rare coincidence.

\section{Discussion}
\label{discu}
In the standard view, the rotating single WD derive from the rotating cores of
giants, which somehow avoided spinning down in the slowly rotating convective
envelope. I argue, instead, that rotating cores in giants are an unattractive idea, especially if these cores are magnetic. Unless the magnetic WD acquired their fields {\it after} emerging from the envelope, the observed dipole moments are so large that a strong interaction with the slowly rotating convective envelope would be very hard to avoid.

I recall the classical demonstration (e.g. in Mestel 1953, 1961) that
rather weak magnetic fields (magnetic energy a small fraction of the rotational
energy) can already transmit enough torques to maintain corotation between core and envelope. Such a weak field could be inherited from the star formation process. In order to prevent these torques from acting, any magnetic field in the core would have to be very weak or very accurately shielded from the convective envelope.
In addition, a differentially rotating, initially nonmagnetic core is unstable to the growth of a small scale dynamo magnetic field, initiated by a magnetic shear instability (Balbus and Hawley 1992). The conditions for existence of this instability in stars were studied in detail already by Acheson (1978) who showed, in particular, that thermal diffusion allows it to operate under a much wider range of conditions than in the adiabatic case.

The very weak differential rotation in the core of the Sun (e.g. Kosovichev et al.1997), for which no good explanation has been put forward except magnetic torques, is strong evidence for the operation of magnetic effects. While the arguments given here do not constitute a proof, I feel they are sufficiently compelling that approximately uniform rotation is a reasonable hypothesis, and is at least as plausible as the traditional assumption, which implies a core rotating $10^4$--$10^5$ times faster than the envelope for the entire duration of the RGB and AGB.

I have explored the consequences of the assumption of approximately uniform rotation for AGBs stars in the process of shedding their envelopes. If this mass loss is strictly axisymmetric, the remaining core rotates very slowly (period more than 10 years). This is just the consequence of angular momentum conservation: the wind takes away almost the entire envelope, but the specific angular momentum it carries away is that of the stellar photosphere, which is larger than the average specific angular momentum of the envelope.

On the other hand, only small nonaxisymmetries in the mass loss process suffice to give the star enough `kick' to explain the angular momentum of single white dwarfs. Such kicks could be associated with mass loss events at the pulsation period of the star or dust-formation episodes in the atmosphere. I have illustrated this with a calculation of the evolution of the probability distribution of the star's angular momentum under the combined action of many small nonaxisymmetric kicks and the angular momentum loss in the wind. The degree of asymmetry required is found to be of the order $10^{-3}$.

Present theories for AGB mass loss are not detailed enough to calculate such asymmetries, but observational indications for asymmetries exist.
Interferometric images of red supergiants ($\alpha$~Sco, $\alpha$~Ori and $\alpha$~ Her: Tuthill, Haniff and Baldwin 1997), speckle reconstructions ($\alpha$~Ori: Kluckers et al.\ 1997) and HST imaging ($\alpha$~Ori: Gilliland and Dupree, 1996) show pronounced `hot spots' on their surfaces. Assuming that such nonuniform photospheric conditions persist during the superwind phase, one would expect them to also affect the dust formation that is thought to be essential for the driving of the wind. The required asymmetry is obtained if a few (5 say) such spots are present, and the wind locally generated above these spots is slightly non-radial by a few tenths of a degree. That the mass flow is indeed asymmetric already close to the stellar photosphere is shown by speckle imaging (IRC 10216: Osterbart et al. 1997), and especially by mm-wave interferometric images of the SiO maser emission. These show a highly clumpy and time dependent structure (Diamond et al. 1994, Humphreys et al. 1996, Pijpers et al. 1994). This maser emission occurs at a few stellar radii, which is also the region where the backreaction of the wind on the star (`kick') takes place. Though the SiO maser emission is very sensitive to small changes in the local physical conditions, models of the emission (Lockett and Elitzur 1992, Bujarrabal 1994) should give estimates of the degree of inhomogeneity in the physical conditions in the wind.

Measurement of deviations from radial flow in proper motion studies of the masing clumps in the wind should enable direct determination of the asymmetries relevant for the kick process described in this paper.

An issue mentioned here only briefly is the origin of the 5 or so very slowly rotating ($P\gapprox 100$yr) magnetic white dwarfs. A possible explanation is angular momentum loss in a radiation driven, but magnetized, wind during post-AGB evolution. This possibility will be further explored elsewhere.

The coupling between core and envelope proposed here would also imply that the cores
of pre-supernovae on the giant branch are so slowly rotating that very
slowly ($P\sim 1$hr) rotating neutron stars would result even if angular momentum were conserved during core collapse. While these would not show up as
pulsars, one would have to argue that none of the observed pulsars were formed in red giants, which feels like an unattractive idea. It turns out, however, that the kicks neutron stars receive at birth and which give them their high observed space motion, are strong enough to impart a significant rotation as well. This idea is developed further in a separate paper (Spruit and Phinney, 1998).

\begin{acknowledgements}
The author thanks Sterl Phinney for many stimulating discussions, and for talking
him out of binary scenarios for rotating white dwarfs. He thanks Martin Groenewegen for discussions on the phenomenology of mass loss of AGB stars, Stan Owocki for pointing out an error in original the calculations, and Frank Pijpers for detailed comments on an earlier version of the text.

\end{acknowledgements}


\begin{thebibliography}{}

\bibitem{}
Acheson D.J., 1978, Phil Trans. Roy. Soc. London {\bf A}., 289, 459

\bibitem{}
Acheson D.J., 1979, Solar Phys. 62, 23

\bibitem{}
Balbus S., Hawley J.F., 1992, ApJ 376, 214

\bibitem{}
Balbus S.A., Hawley J.F, 1994, MNRAS 266, 769


\bibitem{}
Barvainis R., McIntosh G., Predmore C.R., 1987, Nature 329, 613 

\bibitem{}
Becker, R., 1978, Theorie der W\"arme, Springer, Berlin, p.286

\bibitem{}
Bergeron P., Fontaine G., Brassard P. et al. 1993, A.J. 106, 1987

\bibitem{}
Bl\"ocker T., Sch\"onberner D., 1997, A\&A 324, 991

\bibitem{}
Brandenburg A., Nordlund {\AA}., Stein R.F., Torkelsson U., 1995, ApJ 446, 741

\bibitem{}
Bujarrabal V., 1994, A\&A 285, 971

\bibitem{}
Chaplin W.J., Elsworth, Y., Howe R. et al., 1996, NNRAS 280, 849

\bibitem{}
Charbonneau P., MacGregor K.B., 1993, ApJ 430,387

\bibitem{}
Clemens J.C., 1994, PASP 106, 1322

\bibitem{}
Corbard T., Bertomieu G., Morel P. et al., A\&A 324 298

\bibitem{}
Dehner B.T., Kawaler S.D., 1995, ApJ 445, L141

\bibitem{}
Diamond P.J., Kemball A.J., Junor W. et al., 1994, ApJ
430, L61

\bibitem{}
Elsworth, Y., Howe R., Isaak G.R. et al.,
1995, Nature 376,669

\bibitem{}
Fleischer A.J., Gauger A., Sedlmayr E., 1992, A\&A 266, 321

\bibitem{}
Frank A., Mellema G., 1994, ApJ 430, 800

\bibitem{}
Fricke K., 1969, A\&A 1, 388

\bibitem{}
Gilliland R.L., Dupree A.K., 1996, ApJ 463, L29

\bibitem{}
Goldreich P. Nicholson P.D., 1989, ApJ 342, 1075

\bibitem{}
Habing H.J., 1990, in From Miras to Planetary Nebulae, eds M.O. Menessier, A. Omont, Paris: Editions Fronti\`eres, p.19

\bibitem{}
Han Z., Podsiadlowski P., Eggleton P., 1995, MNRAS 272, 800

\bibitem{}
Hawley J.F., Gammie C.F., Balbus S.A., 1995 ApJ 440, 742

\bibitem{}
Heber U., Napiwotzki R., Reid I.N., 1997, A\&A 323, 819

\bibitem{}
H\"ofner S., Dorfi E.A., 1997, A\&A 319, 648

\bibitem{}
Humphreys E.M., Gray M.D., Yates J.A., et al., 1996, MNRAS 282, 1259

\bibitem{}
Icke V., Balick B., Frank A., 1992, A\&A 253,  224

\bibitem{}
Kato S., 1992, PASJ 44, L31

\bibitem{}
Kawaler S.D., O'Brien M.S., Clemens J.C., 1995, ApJ 450,350

\bibitem{}
Kemball A.J., Diamond P.J., 1997, ApJ 481 L111  

\bibitem{}
Kepler S.O., Giovannini O., Wood M.A. et al., 1995, ApJ 447, 874

\bibitem{}
Kluckers, V.A., Edmunds M.G., Morris R.H. et al.,1997, MNRAS 284, 711

\bibitem{}
Koester, D., Herrero, A., 1988, ApJ 332, 910


\bibitem{}
Kolb U., 1993 A\&A 271, 149

\bibitem{}
Kosovichev A.G., Schou  J., Scherrer P.H., et al., 1997, SP 170, 43



\bibitem{}
Liebert J., 1995, in D.A.H. Buckley, B. Warner, eds., Cape workshop on magnetic
cataclysmic variables, ASP conf. series 85, p59.

\bibitem{}
Livio M., 1996, in Evolutionary Processes in Binary Stars, eds. R. Wijers, M.B.
Davies, C.A. Tout (NATO ASI series C 477), Kluwer, Dordrecht, p.141

\bibitem{}
Lockett P., Elitzur M., 1992, ApJ 399, 704

\bibitem{}
Matsumoto R., Tajima T., 1995, ApJ 445, 767

\bibitem{}
Mestel L., 1953, MNRAS 113, 716 (remark on p. 735)

\bibitem{}
Mestel L., 1961, MNRAS 122, 473



\bibitem{}
Nedoluha G.E., Watson W.D., 1994, ApJ 423, 394 

\bibitem{}
O'Brien M.S., Clemens J.C., Kawaler S.D. et al., 1996, ApJ 467, 397

\bibitem{}
Osterbart R., Balega Y.Y., Weigelt G. et al., 1997, in Planetary Nebulae, ed.\ H.\ Habing, (IAU Symp 180), in press

\bibitem{}
Perinotto M., 1990, in Angular momentum transport and mass loss from hot stars, eds. L.A.\ Willson, R.\ Stalio (NATA ASI C316), Kluwer Dordrecht, p.291

\bibitem{}
Pfeiffer B., Vauclair G., Dolez N. et al., 1995, Baltic Astron. 4, 245

\bibitem{}
Pijpers F.P., 1993, in Mass loss on the AGB and beyond, ed. H.E. Schwarz, (Proc 2${\rm nd}$ ESO-CTIO workshop), ESO, Garching, p.26

\bibitem{}
Pijpers F.P., Hearn A.G., 1989, A\&A 209,198

\bibitem{}
Pijpers F.P., Pardo J.R., Bujarrabal V., 1994, A\&A 286, 501

\bibitem{}
Press W.H., 1981, ApJ 245, 286

\bibitem{}
Putney A., 1996, PASP 108, 638



\bibitem{}
Reid, I.N., 1996, PASJ  111, 2000

\bibitem{}
Sakurai Takeo, Kitayama O., Ma J., 1995, Geophys. Astrophys. Fluid Dyn. 79, 277

\bibitem{}
Schatzman E., 1993, A\&A 279, 431

\bibitem{}
Schmidt G.D., Norsworthy J.E., 1991, ApJ 366, 270

\bibitem{}
Schmidt G.D. \& Smith P.S., 1995, ApJ 448, 305

\bibitem{}
Schmidt G.D., West S.C., Liebert J. et al., 1986, ApJ 309, 218

\bibitem{}
Sedlmayr E., Carsten D., 1995, Space Sci. Rev. 73, 211

\bibitem{}
Sch\"onberner D., 1990 in From Miras to Planetary Nebulae, eds. M.O. Menessier, A. Omont, Editions Fronti\`eres, p. 355

\bibitem{}
Sion E.M., Fritz M.L., McMullin J.P. et al., 1988, AJ 96, 251

\bibitem{}
Spruit H.C., 1987, in The internal solar angular velocity, eds. B.R. Durney, S. Sofia, Reidel, Dordrecht, p.185

\bibitem{}
Spruit H.C., Phinney E.S., 1998, Nature, submitted

\bibitem{}
Spruit H.C., Knobloch E., Roxburgh I.W., 1983, Nature   304, 520

\bibitem{}
Talon S., Zahn J.-P., 1998, A\&A  329, 315


\bibitem{}
Taam R.E., 1995, in J. van Paradijs et al., eds., Compact stars in binaries, (IAU
Symp 165), Kluwer, Dordrecht p.3

\bibitem{}
Tomczyk S., Schou J., Thompson M.J., 1996, Bull. Astron. Soc. India 24, 245

\bibitem{}
Tuthill P.G., Haniff C.A., Baldwin J.E., 1997, MNRAS 285, 529

\bibitem{}
Urpin V.A., 1996 MNRAS 280,149

\bibitem{}
Vassiliadis E., Wood P.R., 1993, ApJ 413, 641

\bibitem{}
Verbunt F., 1996, in Evolutionary Processes in Binary Stars, eds. R. Wijers,
M.B. Davies, C.A. Tout (NATO ASI series C 477), Kluwer, Dordrecht, p. 201


\bibitem{}
Wesemael F., Auer L.H., Van Horn H.M. et al., 1980, ApJS 43, 159

\bibitem{}
Winget D.E., Nather R.E., Clemens J.C. et al., 1991, ApJ 378, 325

\bibitem{}
Winget D.E., Nather R.E., Clemens J.C. et al., 1994, ApJ 430, 839

\bibitem{}
Zahn J.-P., 1974, in Stellar Instability and Evolution, eds. P.Ledoux et al., Reidel, Dordrecht, p.185

\bibitem{}
Zahn J.-P., 1990, in Inside the Sun, eds. G. Berthomieu, M. Cribier,
Kluwer, Dordrecht, p.425

\bibitem{}
Zahn, J.-P., Talon S., Matias J., 1997, A\&A 322 320

\end{thebibliography}
\end{document}